\documentclass{ws-mpla}

 \usepackage{hyperref}
 \usepackage[super]{cite}
 \usepackage[utf8]{inputenc}
 \usepackage{amsmath,amssymb}
 \usepackage[english]{babel}
\usepackage{wrapfig}
 \usepackage{graphicx}
\usepackage{epstopdf}
 
\def\URL#1{\url{<#1>}} 

\begin{document}

\markboth{L.~D.~Kolupaeva, K.~S.~Kuzmin, O.~N.~Petrova, I.~M.~Shandrov}
{Some uncertainties of neutrino oscillation effect in the NO$\nu$A experiment}

\catchline{}{}{}{}{}

\title{Some uncertainties of neutrino oscillation effect\\ in the NO$\nu$A experiment}

\author{\footnotesize LYUDMILA~D.~KOLUPAEVA\/$^{a,b}$, 
                      KONSTANTIN~S.~KUZMIN\/$^{a,c}$,\\ 
                      OLGA~N.~PETROVA\/$^{a}$\footnote{redponick@gmail.com},
                      and IGOR~M.~SHANDROV\/$^{a}$ 
       }
\address{$^{a}$Joint Institute for Nuclear Research, RU-141980 Dubna, Russia,\\
         $^{b}$Moscow State University, RU-119991 Moscow, Russia,\\
         $^{c}$Institute for Theoretical and Experimental Physics, RU-117218 Moscow, Russia}

\maketitle

\pub{Received (26 Nov 2015)}{Revised (8 Jan 2016)}

\allowdisplaybreaks

\begin{abstract}
Uncertainties related to the effect of neutrino coherent forward scattering in Earth's matter (MSW mechanism) and 
with the cross sections of quasi-elastic neutrino scattering on nuclear targets of the NO$\nu$A detectors are studied. 
The NO$\nu$A sensitivity to the neutrino mass hierarchy and the CP violating phase is discussed.
\end{abstract}

\ccode{PACS Nos.: 13.15.+g, 13.60.Hb, 14.60.Lm, 14.60.Pq, 25.30.Pt}

\section{Introduction}

NO$\nu$A is an accelerator experiment at FNAL (USA)~\cite{Patterson:2012zs} devoted for studying neutrino oscillation phenomenon. 
This is one of so-called ``off-axis'' new generation experiments with two detectors sited 14~mrad off the NuMI beam axis and separated 
by 810~km of the Earth stratum. Near and Far Detectors are identical, except for the volume 
(300~ton for the Near Detector and 14~kton for the Far Detector).
The first one is intended for the flux calibration and normalization of neutrino interaction cross sections,
another one is just for the neutrino mixing parameter measuring. 
It is used, as a target, a scintillator consisting of a mixture of mineral oil and pseudocumene that fills up the PVC cell structure. 
The detector design makes it possible to identify the event topologies singling out, with a good precision, 
the quasi-elastic (QE) neutrino scattering on the nuclei.

The major goals of the NO$\nu$A experiment are 
the hierarchy determination in the neutrino mass spectrum,
measurement of the CP violating phase $\delta_{\text{CP}}$, 
determination of the mixing angle $\theta_{23}$ octant, and 
adjustment of the mixing angle $\theta_{13}$.
After two three-year modes of data acquisition with the neutrino (antineutrino) beams, it is expected 68 (32)
signal events caused by flavor transitions $\nu_{\mu}\to\nu_{e}$ ($\bar\nu_{\mu}\to\bar\nu_{e}$), 
and about 500 (270) events from survived $\nu_{\mu}$ ($\bar\nu_{\mu}$)~\cite{Backhouse:2015nga}.

The impact of the CP violating phase and neutrino mass hierarchy to the measured $e^{\pm}$ event rates 
in the Far Detector essentially depends on the density of the matter through which the $\nu_{\mu}$ and $\bar\nu_{\mu}$ beams propagate.

Under conditions of the NO$\nu$A experiment, the matter density varies with distance along the neutrino beam; 
the impact of this effect is studied in the Section~\ref{NonConstantDensity}. It will be shown that it is possible, with sufficient 
accuracy, to use a certain effective value of constant density instead of the realistic (followed from the CRUST~1.0 
model~\cite{CRUST1.0}) density distribution. We also estimate a possible uncertainty of this value. 
Also, an estimate of the uncertainty for this value will be done. 

The obtained effective density is used in Section~\ref{NOvASensitivity} for studying NO$\nu$A sensitivity to the neutrino mass hierarchy 
in accordance with the CP violating phase. For a medium with constant density, probability of flavor transitions $\nu_{\mu}\to\nu_{e}$ 
and $\bar\nu_{\mu}\to\bar\nu_{e}$ can be obtained from the exact solution of evolution equation (see., e.g.\ Ref.~\cite{Naumov:1991rh}). 
However, approximate formulas~\cite{Cervera:2000kp,Minakata:2012ue} are often used to simplify calculations.
The permissibility of these approximations for the sensitivity estimation is studied in the same section
\footnote{For other cases, in order to calculate the probability of flavor oscillations in matter, the exact formulas are being applied.
	  In all subsequent calculations, it is used the set of mixing angles and neutrino mass-squared splittings recommended
          in the review~\cite{Agashe:2014kda}.
         }.

Section~\ref{CrossSections} is devoted to an empirical description of the QE scattering cross sections of (anti)neutrino on nuclear targets, by applying the method of running nucleon axial mass $M_A^{\text{run}}$ within the framework of the Relativistic Fermi Gas (RFG) 
model~\cite{Smith:1972xh}. The method allows us to phenomenologically account for the nuclear effects that are not described by the RFG 
model, but significant in the NO$\nu$A energy range. 
The impact of the corresponding modification of the QE cross sections to the expected event rates in the Far Detector is studied.

\section{Nonconstant Matter Density Effect}
\label{NonConstantDensity}

Sensitivity of the NO$\nu$A experiment to the mass hierarchy and CP violating phase is defined by the effect of neutrino coherent forward
scattering on electrons when the beam propagates through the matter of Earth~\cite {Wolfenstein:1977ue,Mikheev:1986gs,Mikheev:1986wj}.
According to geological measurements, the average density of continental crust in North America ranges approximately from 2.7 to 
3.0~g/$\text{cm}^{3}$~\cite{Hasterok}.
In practice, to simplify the analysis, a constant density is used instead of the density varying along the beam trajectory.
The feasibility of this approach for the experiment NO$\nu$A is demonstrated below, and the effective density value is defined,
which best way reproduces the results obtained with the model density profile. 

\begin{wrapfigure}{R}{0.65\textwidth}
\includegraphics[width=0.6\textwidth]{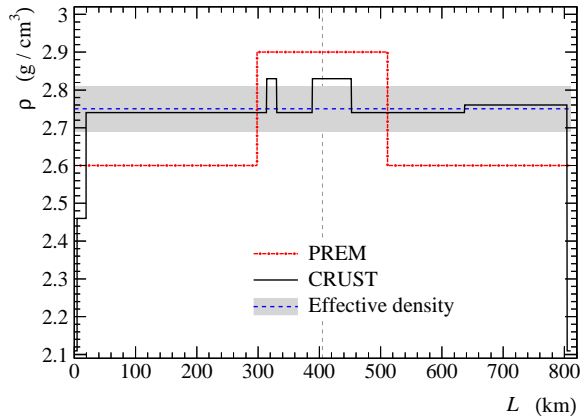}
\caption{\label{Fig1}The matter density distribution in Earth's continental crust on the neutrino path from the source 
         to the Far Detector of the NO$\nu$A experiment, calculated according to PREM~\cite{Dziewonski:1981xy} 
         (dot-dashed line) and CRUST~\cite{CRUST1.0} (solid line). The effective density~\eqref{rho_eff} 
         (dashed line with shaded area) is shown.}
\end{wrapfigure}

To describe the realistic matter density distribution, the modern model of Earth's crust, CRUST~1.0~\cite{CRUST1.0} (see~Fig.~\ref{Fig1})
has been chosen. 
This model, covered whole Earth, is based on the newest seismological data, including these on transverse and longitudinal 
seismic wave velocities, and involves information on the depths of crustal boundaries and matter densities in all crustal layers,
represented on a coordinate grid with averaging out $1^\circ\!\times1^\circ$.

For the effective constant density, $\tilde{\rho}$, we propose to use the value providing the minimum root-mean-square deviation
of the expected rate of $e^\pm$ events in the Far Detector in comparison with the result of calculations performed with the variable model
density $\rho(l)$, where $l$ is the distance from the neutrino source, by taking into account the uncertainty of $\rho$ for each piece
of the path in which the value of $\rho$ remains constant.

Ref.~\cite{CRUST1.0} does not provide information on the uncertainties in the density values. In the earlier
and less detailed model PREM~\cite{Dziewonski:1981xy}, a typical uncertainty in the mean density value for a 100~km thickness layer 
is $\sim5$\%~\cite{Gu:2005er}. While the accuracy of the density estimation in the CRUST~1.0 model exceeds, as it should be expected,
that of the PREM, we assume, for a conservative estimation, that the maximum uncertainty $\Delta\rho$ is $\pm 5$\%
in each constant density region.

By spacing the beam trajectory onto the parts ${\Delta}l_j$ we generate, for each $j$, a sample of $N=5000$ density values,
$\rho^\prime_{ij}$ ($i=1 \ldots N$), deviating from $\rho(l\in{\Delta}l_j)$ according to the normal distribution law.
By this means we obtain a series of $N$ density profiles defined by a piecewise function $\rho^\prime_i(l\in \Delta l_j)=\rho^\prime_{ij}$.
A certain value of constant effective density $\tilde\rho_{i}$ can be assigned to each of such profiles, according to the above definition.
The lengths of the constant density regions, ${\Delta}l_j\sim100$~km, were chosen according to the data representation structure in the
CRUST~1.0 model, by taking into account, in particular, that splitting up of the trajectory onto smaller parts would be in excess of 
precision, since the linear size of the $1^\circ\times1^\circ$ pixel is approximately 111~km.

\begin{figure*}[ht]
\includegraphics[width=\textwidth]{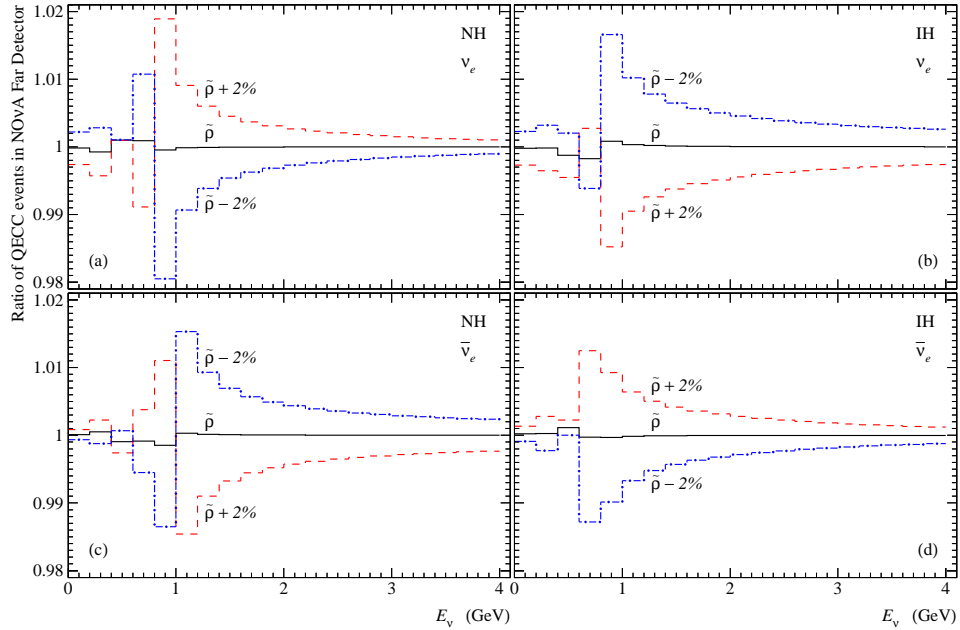}
\caption{\label{Fig2}The ratio of the quasi-elastic event rates induced by electron neutrinos (a,b) 
         and antineutrinos (c,d) in the NO$\nu$A Far Detector,
         calculated by using the effective density~\eqref{rho_eff} and variable density according the CRUST~1.0 model, 
         for the cases of normal hierarchy (NH) (a,c) and inverse hierarchy (IH) (b,d) of neutrino masses.}
\end{figure*}

As a result, for the required effective density $\tilde{\rho}$ we obtain
\begin{equation}
\label{rho_eff}
\tilde{\rho}=\frac{1}{N}\sum\limits_{i=1}^N \tilde\rho_{i}=2.75 \pm 0.06~\text{g}/\text{cm}^{3},
\end{equation}
where the shown error has been calculated as the standard deviation of the arithmetic mean. The resulting effective density $\tilde\rho$
and its uncertainty are shown in Fig.~\ref{Fig1}, together with the density profiles of the CRUST~1.0 and PREM models.
Figure~\ref{Fig2} shows the ratio of the QE event rates calculated with the fixed values of the effective density of the
continental crust~\eqref{rho_eff}, to the event rates obtained by using the variable density according to the CRUST~1.0 model. 
In the calculation of the QE interaction cross section of (anti)neutrinos with the NO$\nu$A detector target,
the RFG model by Smith and Moniz~\cite{Smith:1972xh} (in the version described in Ref.~\cite{Kuzmin:2007kr}) has been used.
To test the stability of the result, similar calculations were performed with taking into account the inelastic processes
(single pion neutrinoproduction, deep inelastic scattering); in these calculations, the results of Refs.~\cite{Kuzmin:2006dh,Kuzmin:2006dt}
were used. The obtained value of $\tilde{\rho}$ remained  practically unchanged. In other words, the value of $\tilde{\rho}$ is
insensitive to the contributions of inelastic reactions. Moreover, as the analysis showed, it is also insensitive to variations of
the nucleon axial mass within a wide range (see Section~\ref{CrossSections}) and of the CP violating phase, $\delta_{\text{CP}}$, in the range of $(0,2\pi)$.

\section{Sensitivity to Neutrino Mass Hierarchy}
\label{NOvASensitivity}

In order to estimate NO$\nu$A sensitivity to neutrino mass hierarchy we have used open-source software package GLoBES
(General Long Baseline Experiment Simulator)~\cite{Huber:2004ka} which allows to generate data for an abstract user-defined
neutrino experiment. Oscillation parameters and some additional information which allows to estimate potential of the
experiment, can be extracted from event energy spectrum generated 
by this package.
The main GLoBES tool is the standard $\chi^2$ function calculation for quantities obeyed Poisson distribution:
\begin{gather*}
\chi^2  = 2 \sum_i \Lambda_i+\chi^2_{\text{prior}}+\chi^2_{\text{pull}}, \quad
\Lambda_i = N_i^{\text{th}}-N_i^{\text{obs}}\left(1-\ln\frac{N_i^{\text{obs}}}{N_i^{\text{th}}}\right), \\
N_i^{\text{obs}} = \left(1+a+br_i\right)s_i+\left(1+c+dr_i\right)b_i, \\
r_i = \frac{E_{i} - E_{\text{mean}}}{E_{\text{max}} - E_{\text{min}}}, \quad
\chi^2_{\text{pull}} = \frac{a^2}{\sigma_a^2}+\frac{b^2}{\sigma_b^2}+\frac{c^2}{\sigma_c^2}+\frac{d^2}{\sigma_d^2},
\end{gather*}
where $i$ is number of neutrino energy bin ($E_{i}$), $N_i^{\text{th}}$ ($N_i^{\text{obs}}$) is predicted (observed) number of events per bin;
$E_{\text{min}}$, $E_{\text{max}}$, and $E_{\text{mean}}$ are, respectively, minimal, maximal and mean neutrino energy in the experiment;
$s_i$ ($b_i$) is number of signal (background) events;
$a$ ($c$) and $b$ ($d$) are values associated with spectrum normalization and calibration for the signal (background);
$\chi^2_{\text{prior}}$ is oscillation parameters priori error contribution, except $\delta_{\text{CP}}$;
$\chi^2_{\text{pull}}$ is sum of the systematical error contributions, determined by pull-method~\cite{Fogli:2002pt}.

The event rate is defined as a product of neutrino flux, oscillation probability, cross sections of neutrino interaction 
with detector material, and detection efficiency for the events of certain type.
The experiment has been determined in the GLoBES configuration file according to its declared characteristics.
Effective density~\eqref{rho_eff}, obtained in the Section~\ref{NonConstantDensity}, has been used on the whole trajectory of neutrino beam.
Proton beam power has been taken to be $0.7$~MW, with its intensity $6\times10^{20}$~POT (protons on target) per year.
The Far Detector characteristics declared in Ref.~\cite{Agarwalla:2012bv} have been used in the calculations.

For the mass hierarchy sensitivity estimation following quantity is used
\begin{equation}
\label{Sensitivity}
\sqrt{\Delta \chi^2} = \sqrt{\left|\chi^2_{\text{test}}-\chi^2_{\text{true}}\right|}.
\end{equation}
As it was shown in Ref.~\cite{Qian:2012zn}, in general case (and specifically in the problem of the mass hierarchy determining)
the quantity~\eqref{Sensitivity} is not true experiment sensitivity, measured in terms of standard deviations;
however, the estimation~\eqref{Sensitivity} is applicable for studying of relative sensitivity dependence on the mixing parameter variations 
and impact of various approximations.
We consider two cases below:
(a) the true hierarchy (that has been used in calculations of $\chi^2_{\text{true}}$) is normal, 
while the one assumed in the analysis (that has been used in the calculations of $\chi^2_{\text{test}}$) is inverse and
(b) the true hierarchy is inverse, while the one assumed in the analysis is normal.

Approximate expressions for oscillation probability have some advantages over the exact formula, namely, retaining 
sufficient accuracy, they reduce computing time expenses. Because of this, in the data processing of some oscillation experiments 
(see., e.g., Refs.~\cite{Bian:2013saa,Adamson:2014vgd}) that requires, as a rule, very large amounts of computations, just the approximate 
formulas are sometimes used. They are obtained by expansion of the exact solution in a series in the small parameter
$\alpha={\Delta m^2_{12}}/{\Delta m^2_{13}}$ ($|\alpha|\simeq0.03$) with an accuracy of $\mathcal{O}(\alpha)$
or $\mathcal{O}(\alpha^2)$ (in the first case the last term should be dropped)~\cite{Cervera:2000kp,Minakata:2012ue}:
\begin{equation}
\label{ApproximateSolution}
\begin{gathered}
\begin{aligned}
P_{\stackrel{(-)}{\nu_\mu} \to \stackrel{(-)}{\nu_e}}\approx\sin^2{\theta_{23}} \sin^2{2\theta_{13}} \xi^2+
                       \alpha J \cos{(\Delta \pm \delta_{\text{CP}})} \zeta \xi+\alpha^2 \cos^2{\theta_{23}} \sin^2{2\theta_{12}} \zeta^2;
\end{aligned} \\
J = \cos{\theta_{13}} \sin{2\theta_{13}} \sin{2\theta_{12}} \sin{2\theta_{23}}, \\
\zeta=\frac{\sin{\Delta A}}{A}, \quad \xi=\frac{\sin{\Delta(1-A)}}{(1-A)}, \quad
A = \pm \frac{G_F n_e L}{\sqrt{2} \Delta}, \quad
\Delta = \frac{\Delta m_{13}^2L}{4E_\nu}.
\end{gathered}
\end{equation}
Here $G_F$ is the Fermi constant, $n_e$ is electron density, $L$ is the distance between the source and detector,
and $E_\nu$ is neutrino energy. Signs ``$-$'' correspond to antineutrino case.

\begin{figure*}[ht]
\includegraphics[width=0.98\textwidth]{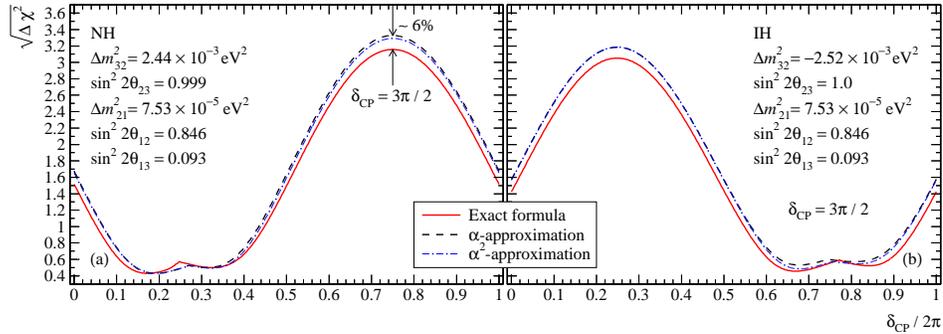}
\caption{\label{Fig3}The experiment sensitivity~\eqref{Sensitivity} to the neutrino mass hierarchy: normal (a) and inverse (b), depending on 
         $\delta_\text{CP}$. Solid curves correspond to the calculation by using the exact formulas~\cite{Naumov:1991rh}, 
         dashed curves are calculated according to Eq.~\eqref{ApproximateSolution} with 
         accuracy of $\mathcal{O}(\alpha)$, dot-dashed curves -- with accuracy of $\mathcal{O}(\alpha^2)$.}
\end{figure*}

Figure~\ref{Fig3} shows the experiment sensitivities to the neutrino mass hierarchy (in dependence on CP violating phase) 
calculated with the exact formula~\cite{Naumov:1991rh} and Eq.~\eqref{ApproximateSolution} in the first and second orders
in $\alpha$; the $N_{\text{obs}}$ in $\chi^2$ are always calculated with the exact formula.
As it is seen, there is some difference between calculations with the exact and approximate formulas%
\footnote{Approximate formula curves dispose closer to each other rather than to exact formula curve due to the procedure of $\chi^2$ function minimization. It choose different values of oscillation parameters for each of these curves to reduce difference between observed and predicted number of events. This difference exists due to wrong hierarchy and approximate probability formula in fit.}. In particular, for $\delta_{\text{CP}}={3\pi}/{2}$
(the best-fit value for reactor data and recent T2K result~\cite{Abe:2015awa}, also consistent with NO$\nu$A preliminary data
analysis~\cite{Sachdev:2015hpa}) the systematic error associated with usage of the approximate formulas reaches almost 6\% in the normal hierarchy case.
The statistical error for this value of $\delta_{\text {CP}}$ is expected to be $\sim10\%$. 

\section{Uncertainties related to quasi-elastic neutrino scattering on nuclei}
\label{CrossSections}

Uncertainty in description of the cross sections for the (anti)neutrino scattering on detector target nuclei 
affects the accuracy of extraction of the neutrino mixing parameters, such as $\Delta{m_{13}^2}$ and 
$\theta_{23}$ (see, for example, Refs.~\cite{Meloni:2012fq,Coloma:2013tba,Benhar:2015wva}), $\theta_{13}$ and 
$\delta_\text{CP}$~\cite{FernandezMartinez:2010dm,Benhar:2015wva}, from the oscillation experiments.
The largest uncertainty in the QE cross section calculation is brought by the nucleon axial form factor $F_{A}(q^2)$ 
(see, e.g., Ref.~\cite{Kuzmin:2007kr}). In the conventional dipole parametrization for the dependence of $F_{A}$
on the 4-momentum  transfer square, $q^2$, this uncertainty reduces to the error of the parameter $M_{A}$ (so-called nucleon axial mass),
extracted from the data on QE (anti)neutrino scattering on nuclei.

The analysis of the experimental data on the $\bar\nu_{\mu}\text{H}$ and $\nu_{\mu}\text{D}$ QE scattering (total and differential 
cross sections, $q^2$-distributions)~\cite{hypKN} yields
\[
M_{A} = 1.003_{-0.084}^{+0.085}~\text{GeV}
\quad
\left({\chi^{2}}/{\text{dof}} = \frac{124.6}{117-7} \approx 1.13\right),
\]
that agrees within the errors with the result of the earlier data processing~\cite{Bodek:2007vi}, $M_A=1.014\pm0.014$~GeV,
as well as with the result of a global analysis of the data obtained before 2007~\cite{Kuzmin:2007kr}. However, new experiments 
with heavier nuclei as targets yield a wide range of values of $M_{A}$ extracted within the RFG model, and tend to have an increase
of $M_{A}$ with decrease in average energy of the neutrino beam, $\langle{}E_{\nu}\rangle$; examples of some recent results are given in the table. 
\begin{table}[h]
\label{expMA}
\centering
\begin{tabular}{c|c|c|c}
\hline
Experiment                             & Nucleus & $\langle{}E_{\nu}\rangle$ (GeV) & $M_{A}$ (GeV)          \\
\hline
NOMAD~\cite{Lyubushkin:2008pe}         &   C     & 24.3 [$\nu_{\mu}$]              & $1.05 \pm 0.06$        \\
                                       &   C     & 17.2 [$\bar\nu_{\mu}$]          & $1.06 \pm 0.14$        \\
MINOS~\cite{Adamson:2014pgc}           &   Fe    & 2.79 [$\nu_{\mu}$]              & $1.23 \pm 0.18$        \\
K2K~\cite{Gran:2006jn}                 &   O     & 1.30 [$\nu_{\mu}$]              & $1.2 \pm 0.12$         \\
T2K~\cite{Abe:2015oar}                 &   C     & 0.86  [$\nu_{\mu}$]             & $1.45^{+0.26}_{-0.30}$ \\
MiniBooNE~\cite{AguilarArevalo:2010zc} &   C     & 0.788 [$\nu_{\mu}$]             & $1.35\pm0.17$          \\
MiniBooNE~\cite{AguilarArevalo:2013hm} &   C     & 0.665 [$\bar\nu_\mu$]           & --                    \\
\hline
\end{tabular}
\end{table}

It should be noted that the experiments listed in the table use, as a rule, somewhat different versions of the RFG model,  
including the value of its parameters, such as the Fermi momenta and binding energies of the nucleons in nuclei%
\footnote{For example, in the MiniBooNE data processing~\cite{AguilarArevalo:2010zc}, an empirical parameter $\kappa$
          which rescales the minimum allowable value of the nucleon energy in the RFG, has been introduced into the model. 
          In the MiniBooNE experiment with the $\bar\nu_\mu$ beam~\cite{AguilarArevalo:2013hm} an extraction of $M_A$
          has not been conducted, but the results are consistent with the calculation performed with $M_A=1.35$~GeV 
          and $\kappa=1.007$ (values obtained from the $\nu_\mu$ beam data~\cite{AguilarArevalo:2010zc}).
         }.

There have been many attempts in recent years to explain the effect of growth of $M_A$ within the framework of more sophisticated
(in comparison with the RFG) models of the (anti)neutrino interaction with the nucleus. For instance, we mention the models based on
the so-called spectral functions~\cite{FernandezMartinez:2010dm,Benhar:2015wva} (the RFG model can be considered as the simplest 
particular case), relativistic mean field approach~\cite{FernandezMartinez:2010dm}, and 
models based on the random phase approximation~\cite{FernandezMartinez:2010dm,Chauhan:2011zz}
and on account of multi-nucleon interactions in the nucleus~\cite{Meloni:2012fq,FernandezMartinez:2010dm,Coloma:2013tba,Benhar:2015wva}.

The authors of Ref.~\cite{hypKN} have proposed a phenomenological recipe for description of the QE scattering 
on nuclear targets, built on the abovementioned observation that the effective axial nucleon mass extracted
(based on the RFG model) from the experimental data increases with a decrease of the neutrino energy, and also on
the fact that, in the most successful models, the QE cross section per nucleon weakly depends on the number of nucleons in the nucleus.
The structure functions, $T_i$, describing the QE scattering on a nucleus depend not only on $q^2$ (as in the case of scattering 
on free nucleon), but separately on the energy transfer $q_0$ and momentum transfer $|\mathbf{q}|$, which, taking into account 
the energy-momentum conservation, reduces to the dependence of $T_i$ on $q^2$ and neutrino energy $E_\nu$.
Hence, it is proposed to reduce the effective accounting for thin (beyond the RFG) nuclear effects to a replacement of the parameter $M_A$ 
in the structure functions $T_i^{\text{RFG}}$ (calculated within the RFG model) by an empirical function $M_{A}^\text{run}$ 
(``running axial mass'') steadily decreasing with energy $E_\nu$.
A detailed statistical analysis has shown that the simplest parameterization of the form
$M_{A}^\text{run}=M_{0}\left(1+E_{0}/E_{\nu}\right)$ 
provides a good description of the available accelerator data on the total, differential and double differential cross sections 
and $q^2$ distributions for the QE interaction with various nuclear targets. The parameter $M_0$ can be identified with the conventional 
axial mass $M_A$ extracted from the deuterium experiments.

Although any realistic description of the neutrino-nucleus interaction should certainly go beyond the RFG model, the running axial mass of the nucleon effectively ``absorbs'' the major part of the rather involved nuclear dynamics missing from the RFG model. While such an approach potentially may have a limited predictive power, it was demonstrated that it works rather well in the whole kinematic regions of all available experimental data~\cite{hypKN}.

From a joint analysis of data on light (hydrogen and deuterium) and heavy nuclear targets the following values have been obtained 
\begin{equation}
\label{M_0E_0}
M_{0}=1.006\pm0.025~\text{GeV},
\quad
E_{0}=0.334_{-0.054}^{+0.058}~\text{GeV}.
\end{equation}
The value of $\chi^{2}/\text{dof}$ following from the analysis is equal to $297.9/(435-19)=0.72$.
The $\nu_\mu$ and $\bar\nu_\mu$ flux normalization factors were considered as free parameters included
into the standard ``penalty terms'' with the errors given by the authors of each individual experiment%
\footnote{The data selection criteria are described in Ref.~\cite{Kuzmin:2007kr}. 
          Also the version of the RFG model used in the analysis is discussed there.}
in all experiments but MINER$\nu$A~\cite{Fields:2013zhk,Fiorentini:2013ezn} and T2K ND280~\cite{Abe:2014iza} 
where the correlations of data errors were taken into account.

\begin{wrapfigure}{R}{0.64\textwidth}
\includegraphics[width=0.59\textwidth]{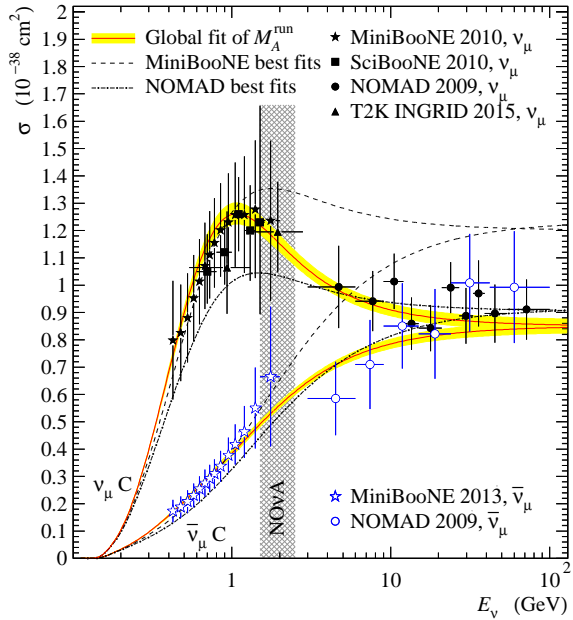}
\caption{\label{Fig4}Total cross sections of quasi-elastic scattering of $\nu_\mu$ and $\bar\nu_\mu$ on a carbon target in comparison 
         with experimental data from NOMAD~\cite{Lyubushkin:2008pe}, SciBooNE~\cite{AlcarazAunion:2009ku,Aunion:2010zz}, 
         MiniBooNE 2010~\cite{AguilarArevalo:2010zc}, 2013~\cite{AguilarArevalo:2013hm}, T2K INGRID~\cite{Abe:2015oar}. 
         Solid curves with narrow bands correspond to calculations with $M_{A}^\text{run}$ and variations of 
         the parameters~\eqref{M_0E_0} within 1$\sigma$. Calculations made with constant values of $M_A$ extracted from the
         MiniBooNE 2010 data are shown by dashed curves, the calculations made with $M_{A}$ from the NOMAD data are shown by dot-dashed
         curves. The shaded area indicates the approximate energy range in the NO$\nu$A experiment.}
\end{wrapfigure}

As an example, Fig.~\ref{Fig4} shows a comparison of the total QE cross sections calculated by using the $M_{A}^\text{run}(E_\nu)$
with the parameters~\eqref{M_0E_0} (solid curves) and data of several experiments with carbonaceous targets
(in terms of pure carbon)~\cite{Lyubushkin:2008pe,AlcarazAunion:2009ku,AguilarArevalo:2010zc,AguilarArevalo:2013hm,Abe:2015oar}.
The narrow bands around the curves indicate the permissible variation of the cross sections due to the uncertainties of the
parameters~\eqref{M_0E_0} within one standard deviation (68\% C.L.).
Also shown are the curves calculated with the constant values of $M_{A}$, extracted from the MiniBooNE~\cite{AguilarArevalo:2010zc} 
and NOMAD~\cite{Lyubushkin:2008pe} data.
The shaded area in Fig.~\ref{Fig4} indicates the approximate energy range of the NO$\nu$A experiment. 
It is seen that the running axial mass successfully compensates for the shortcomings of the RFG model in the low-energy region
and provides a good description of the data at high energies. A comparison of the calculation with the measurements of 
the differential QE cross sections measured by MINER$\nu$A indicates~\cite{hypKN} the applicability of this approach
also to the intermediate energy region, in which the NO$\nu$A experiment operates.

\begin{figure*}[ht]
\includegraphics[width=\textwidth]{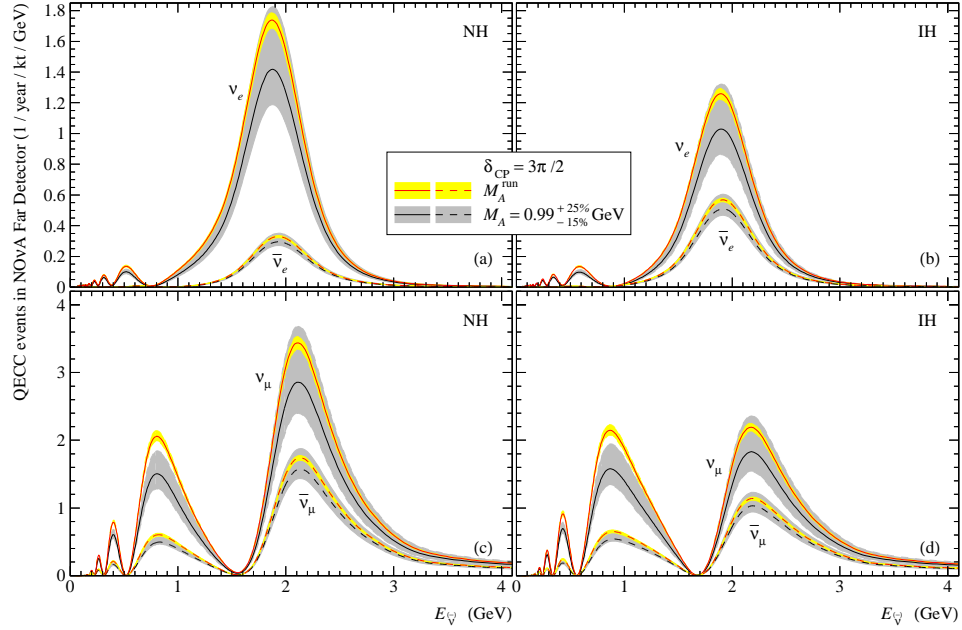}
\caption{\label{Fig5}Rates of quasi-elastic events induced by electron (a,b) and muon (c,d) neutrinos in the $\nu$ mode (solid curves) 
         and antineutrinos in the $\bar\nu$ mode (dashed curves), inside the NO$\nu$A Far Detector, calculated for the normal (a,c) 
         and inverse (b,d) neutrino mass hierarchies (distributions vs. neutrino energy). The calculations are done with 
         the NO$\nu$A analysis default value~\eqref{NOvAdefault} (curves with wide bands) and with $M_{A}^\text{run}$ at $1\sigma$ 
         parameter uncertainties (curves surrounded by narrow bands); $\delta_\text{CP}=3\pi/2$.}
\end{figure*}

\begin{figure*}[ht]
\includegraphics[width=\textwidth]{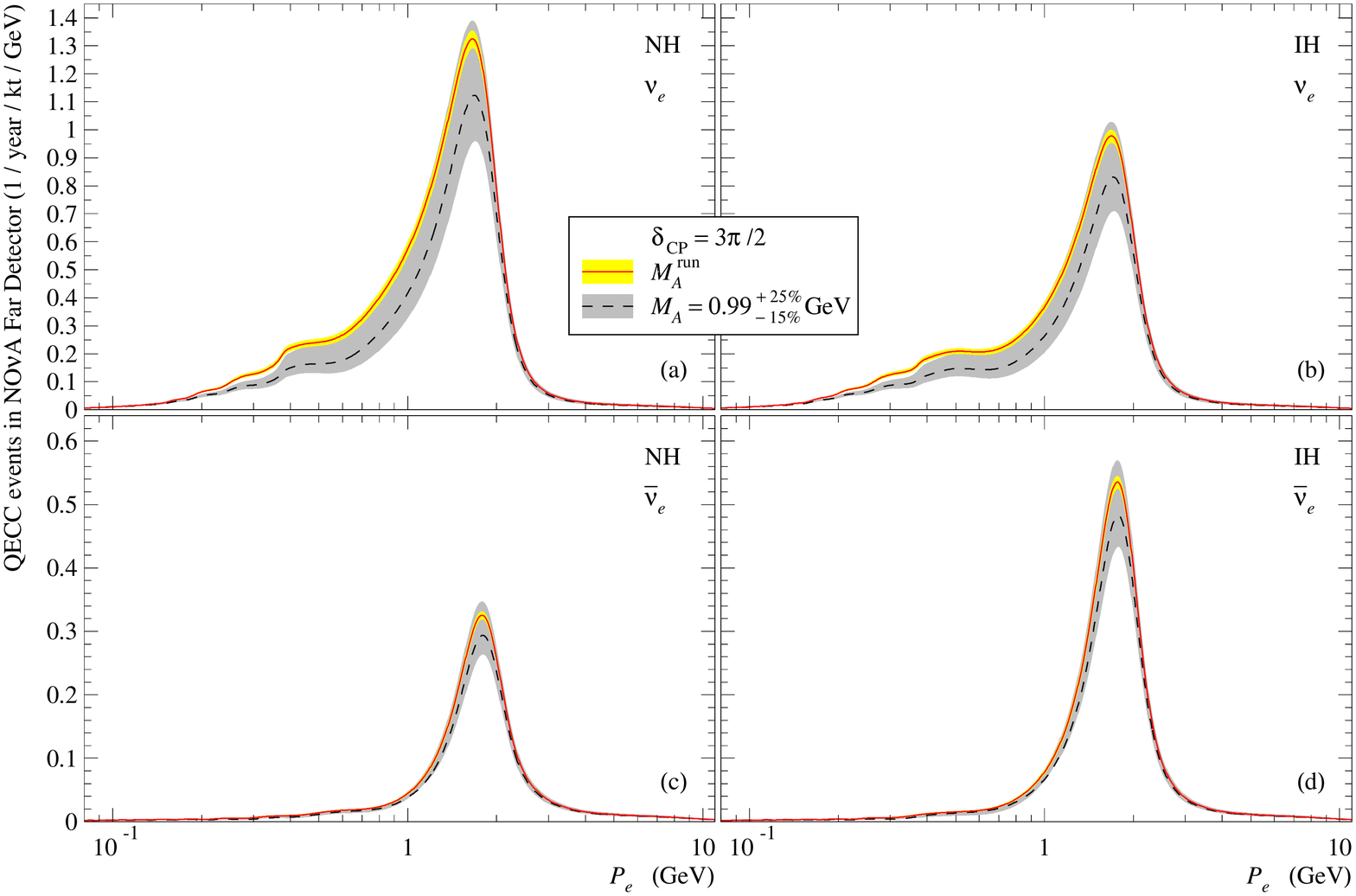}\\
\includegraphics[width=\textwidth]{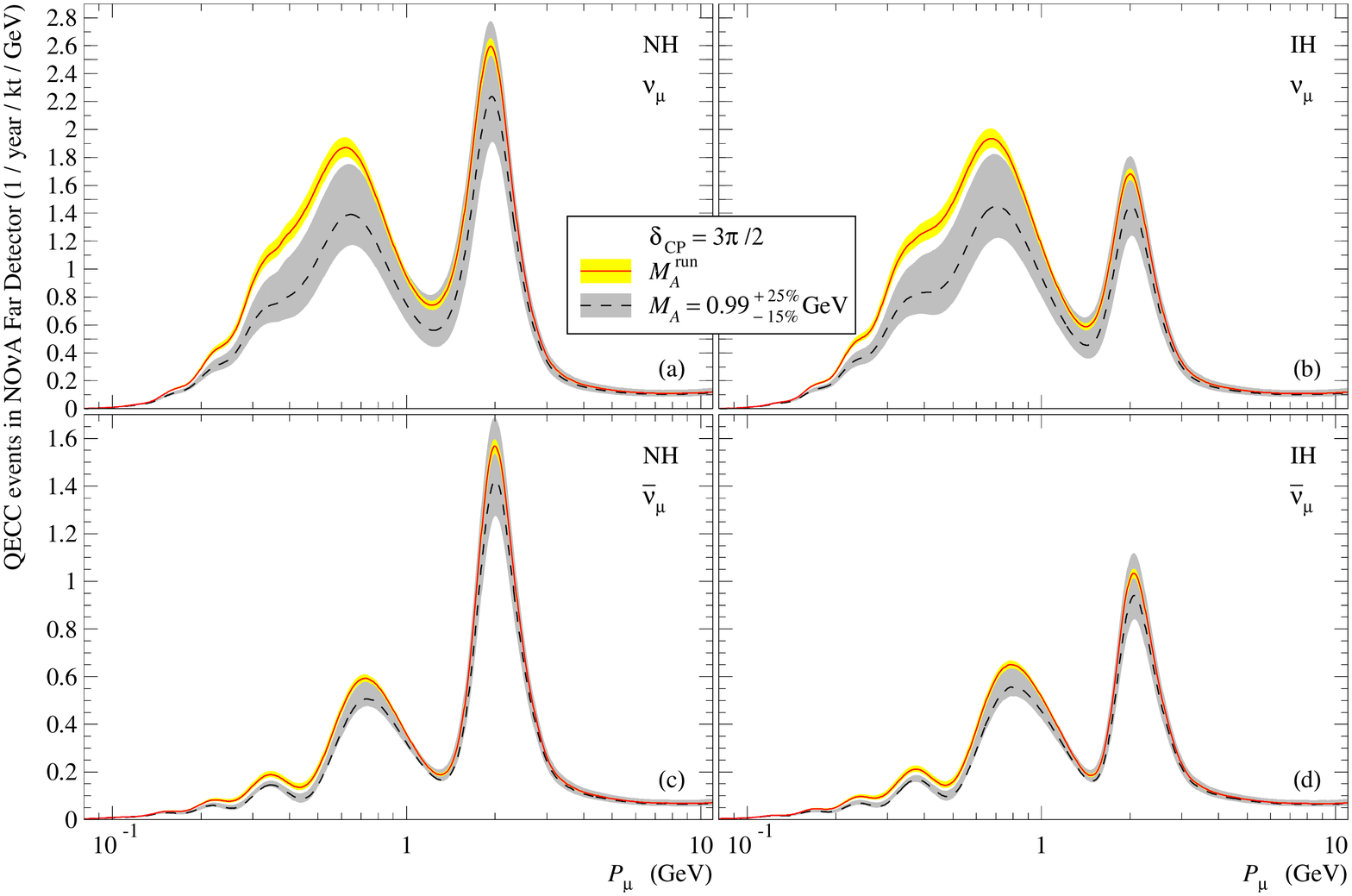}
\caption{\label{Fig6}Rates of quasi-elastic events induced by electron (upper 4 panels) and muon (lower 4 panels) neutrinos 
         in the $\nu$ mode (a,b) and antineutrinos in the $\bar\nu$ mode (c,d), inside the NO$\nu$A Far Detector, 
         calculated for the normal (a,c) and inverse (b,d) neutrino mass hierarchies (distributions vs. charged lepton momentum). 
         Dashed curves with wide bands correspond to the calculations done with the NO$\nu$A analysis default value~\eqref{NOvAdefault}, 
         solid curves with narrow bands correspond to $M_{A}^\text{run}$ calculations at $1\sigma$ parameter uncertainties; 
         $\delta_\text{CP}=3\pi/2$.}
\end{figure*}

In Fig.~\ref{Fig5}, the solid (dashed) curves show the expected $E_\nu$ distributions of the rates of events caused by the QE interactions
of electron and muon neutrinos (antineutrinos) in the NO$\nu$A Far Detector for the $\nu$ ($\bar\nu$) mode of the experiment,
for both normal and inverse neutrino mass hierarchies. The calculations are performed taking into account the flavor transitions
in a matter with $\rho=\tilde{\rho}$, according to Eq.~\eqref{rho_eff}; the same mixing parameters as in the previous sections 
and $\delta_\text{CP}=3\pi/2$ are used. Curves surrounded by narrow bands are the calculations (within the RFG) with the running axial mass 
according to Eq.~\eqref{M_0E_0}, the curves with wide bands of uncertainty are the calculations with
\begin{equation}
\label{NOvAdefault}
M_{A}=0.99_{-0.15}^{+0.25}~\text{GeV}
\end{equation}
(NO$\nu$A default). It is clearly seen that the introducing of the running axial mass leads to an overall increase in the event rates 
and to a significant reduction of uncertainty in the calculations.
Figure~\ref{Fig6} shows expected distributions of the rate of events caused by the QE interactions of electron and muon (anti)neutrinos as in the Fig.~\ref{Fig5} but as function of final lepton momentum. Calculations are done at the same assumptions and inputs as in the Fig.~\ref{Fig5}.
\clearpage

\begin{figure*}[ht]
\includegraphics[width=\textwidth]{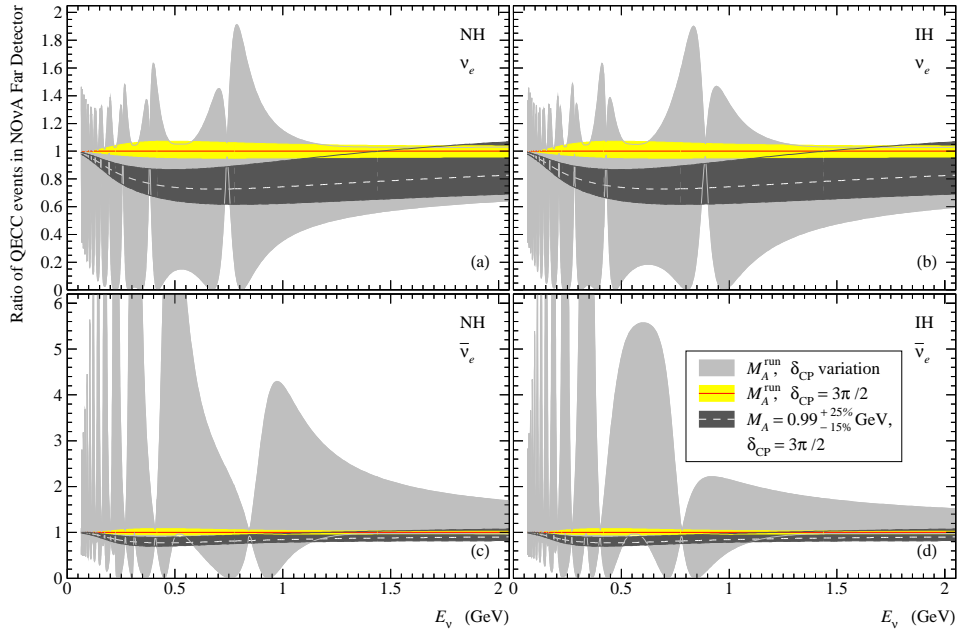}
\caption{\label{Fig7}The relative effects for the rates of quasi-elastic events induced by $\nu_e$ (a,b) and $\bar\nu_e$ (c,d) 
         in the NO$\nu$A Far Detector: 
         the calculations made with constant $M_A$~\eqref{NOvAdefault} and $\delta_\text{CP}=3\pi/2$ (dashed curves with dark bands),
         and with $M_{A}^\text{run}$ for all possible values of $\delta_\text{CP}$ (shaded areas of non-trivial shape) 
         are divided to the calculation made with $M_{A}^\text{run}$ and $\delta_\text{CP}=3\pi/2$ (solid straight lines with narrow 
         light bands) for the cases of normal (a,c) and inverse (b,d) neutrino mass hierarchies.}
\end{figure*}

Figure~\ref{Fig7} shows the relative rates (vs. neutrino energy) of events initiated by the QE interactions of 
the $\nu_e$ and $\bar\nu_e$, occurring in $\nu_\mu$ and $\bar\nu_\mu$ beams as a result of oscillations in Earth's crust. 
The event rates are normalized to a calculation made with the running axial mass, $M_{A}^\text{run}$, 
and with the fixed CP violating phase $\delta_\text{CP}=3\pi/2$. The narrow bands around 
the solid straight lines correspond to the uncertainties of the $M_{A}^\text{run}$ parameters~\eqref{M_0E_0} within 1$\sigma$.
Dark bands around the dashed curves show the calculation made by using Eq.~\eqref{NOvAdefault} and $\delta_\text{CP}=3\pi/2$.
The shaded areas with non-trivial oscillation shape correspond to variation of the $\delta_\text{CP}$ phase within the limits 
from 0 to $2\pi$; the calculation is done with $M_{A}=M_{A}^\text{run}$.

It is seen that the effect of uncertainty in $M_A$ is comparable with the sought effects.
Noteworthy also that in the analysis of the NO$\nu$A experiment, an extrapolation ``Near Detector $\to$ Far Detector'' is used 
and that significantly reduces the impact of this uncertainty to the accuracy of the extracted oscillation parameters.
However, an accurate prediction of absolute rates of events of different types is essential for the correct interpretation
of the measurements and that the accuracy of such predictions is largely dependent on the uncertainties in the
calculation of the cross sections of the QE interactions with nuclei.
\section{Conclusions}

According to our estimation, a variable density of Earth's crust predicted by the CRUST~1.0 model for the NO$\nu$A experiment site 
can be replaced by an effective one with the constant value~\eqref{rho_eff} (close to the average density of the continental crust),
with the difference in the predicted event rates that is beyond the sensitivity of the experiment.

NO$\nu$A experiment sensitivity to the mass hierarchy after full statistics gathering is expected to be $\sim3\sigma$. 
According to our estimation performed with the GLoBES,  for probability measurement 
of flavor transitions $\nu_\mu\longleftrightarrow \nu_e$ and $\bar\nu_\mu\longleftrightarrow\bar\nu_e$ statistical error will reach approximately 10\%
for the maximum CP violation ($\delta_{CP}=3\pi/2$). Error associated with approximate formulas will be about 6\%, so these formulas application to accurate data analysis should be avoided.

The proposed phenomenological description of the QE interactions by the method of running nucleon axial mass, $M_{A}^\text{run}(E_{\nu})$,
can be used for event simulation and data processing in the NO$\nu$A and other experiments studying neutrino oscillations that will 
significantly reduce the systematical error related to the uncertainty of the cross sections of (anti)neutrino interactions with nuclei.

\section*{Acknowledgments}

We would like to thank V.~A.~Naumov, A.~G.~Olshevskiy, O.~B.~Samoylov, D.~V.~Taichenachev, and A.~S.~Sheshukov
for the useful discussions and NO$\nu$A Collaboration for the provided information. The research
has been  supported by the Russian Foundation for Basic Research under Grant No.~14-22-03090 and Grant No.~16-02-01104-a.

\bibliography{references}

\end{document}